\documentclass[nocopyrightspace,preprint,nonatbib]{sigplanconf}
\usepackage[utf8]{inputenc}
\usepackage[T1]{fontenc}
\usepackage{hyperref}
\usepackage{graphicx}
\usepackage{grffile}
\usepackage{longtable}
\usepackage{wrapfig}
\usepackage{rotating}
\usepackage[normalem]{ulem}
\usepackage{amsmath}
\usepackage{textcomp}
\usepackage{amssymb}
\usepackage{hyperref}
\authorinfo{Phil Scott \and Steven Obua \and Jaques Fleuriot}{University of Edinburgh}{}
\copyrightyear{2017}
\usepackage{fancyvrb}
\usepackage{color}
\include{pygments}
\date{\today}
\title{Compiling Purely Functional Structured Programs\thanks{This work has been funded by \href{http://gow.epsrc.ac.uk/NGBOViewGrant.aspx?GrantRef=EP/L011794/1}{EPSRC Grant EP/L011794/1}}}
\hypersetup{
 pdfauthor={Phil Scott},
 pdftitle={Compiling Purely Functional Structured Programs},
 pdfkeywords={},
 pdfsubject={},
 pdfcreator={Emacs 25.1.1 (Org mode 9.0.5)}, 
 pdflang={English}}
\begin{document}

\maketitle
\begin{abstract}
We present a marriage of functional and structured imperative programming that embeds
in pure lambda calculus. We describe how we implement the core of this language in a
monadic DSL which is structurally equivalent to our intended source language and
which, when evaluated, generates pure lambda terms in continuation-passing-style.
\end{abstract}

\section{Pure Functional Programming}
\label{sec:orgc5ba869}
One of the frequently touted benefits of being purely functional lies in the ease
with which one can reason about programs. Pure functions have predictable and
deterministic behaviour. Applying the same pure function to the same argument always
yields the same result anywhere in code, thus allowing us to reason equationally
about a program.

But the lack of global variables and mutable state might be an impractical
restriction for general purpose programming on conventional operating systems, and so
it might be argued that it should be targeted to specific domains, such as the system
configuration language Nix~\cite{NixOS}, where \emph{reproducibility} of system builds is highly
prized.

Our own ProofScript~\cite{ProofScript} language is a new entry into the pure functional programming
space, where the chosen domain is the development of large scale mechanical
verifications of mathematics and software. Our model here is based on massive user
collaboration, with users working within what is effectively a single development
environment.

In this environment, every mechanically verified theorem must be a piece of
reproducible, immutable data, much like a Nix derivation. And so in our programming
language, in which we write the code that produces the verified theorems, we seek the
same discipline of pure functional programming.

\section{Structured Programming}
\label{sec:orgd37f38b}
Pure functional programming typically emphasises recursion and higher order functions
for expressing complex control flow, and this level of abstraction can be
intimidating to those coming from imperative backgrounds and more mainstream
languages such as Python. However, the common imperative idioms from structured
programming are not anathema to the aims of pure functional code. Indeed, the
popularity of \emph{Static Single Assignment}~\cite{SSA} (SSA) in compiler architectures shows that
modelling imperative structured programming in a pure representation is of benefit
when it comes to performing reliable optimisations and static analyses of code.

ProofScript aims for the best of both worlds, marrying structured programming with a
rich functional language whilst maintaining purity. We believe that our language
bears a close affinity to the control flow model assumed in SSA, but with additional
support for first-class functions and mutual recursion.

Syntactically, basic ProofScript is a language with support for lexical scope,
looping and assignment, but with the following key restrictions:

\begin{enumerate}
\item we do not support mutable cells as values;
\item closures close over the values of variables, not their locations;
\item function bodies can only assign to variables whose lexical scope is
contained in the function body.
\end{enumerate}

Rule 1 should not need additional explanation. Being able to create and pass around
mutable storage locations means that the language encodes shared mutable state that
can cross arbitrary regions of code.

Rule 2 rules out situations such as:

\begin{small}
\begin{Verbatim}[commandchars=\\\{\}]
  val x = 0 
  def get () = x
  val a = get ()
  x = x + 1
  val b = get ()
  assert (a == b)
\end{Verbatim}
\end{small}

Here, we have lost the referential transparency of the expression \texttt{get ()}. The
meaning of the function was changed by an imperative update.

Rule 3 rules out common idioms from the world of impure functional programming such
as the imperative (or object-oriented) counter:

\begin{small}
\begin{Verbatim}[commandchars=\\\{\}]
  val counter =
    do
      val i = 0
      def get () = i
      def incr () = i = i + 1
      [get, incr]
\end{Verbatim}
\end{small}

Assignments, therefore, are not expected to interact much with function definitions,
but instead with imperative control flow such as \texttt{while} and \texttt{for} loops, and
conditional \emph{statements}.

\section{CPS}
\label{sec:orge87e7cc}
Before SSA, there was \emph{CPS}, or Continuation Passing Style. In this style, we write
lambda terms where evaluation order has been completely disambiguated so that any
remaining side-effects (such as non-termination) have an explicit ordering.

In CPS, every function takes a future dependent on the function's result. The future,
or continuation, receives this result and outputs the rest of the program, again via
continuation-passing style. For a simple example, consider an ambiguous application
\(f\ x\ (g\ y)\) where all values are atomic. This might be rendered in CPS form by two
different lambda terms

\begin{displaymath}
  \lambda k. g\ y\ (\lambda r. f\ x\ (\lambda r'. r'\ r\ k))
\end{displaymath} 

or

\begin{displaymath}
  \lambda k. f\ x\ (\lambda r. g\ y\ (\lambda r'. r\ r'\ k))
\end{displaymath} 
depending on whether we evaluate the outer application right-to-left or
left-to-right.

Compilation of functional programs via CPS transformation was the basis for the
classic text \emph{Compiling with Continuations}~\cite{CC}, but the approach has
notoriously fallen out of favour. Compiler toolchains such as LLVM~\cite{LLVM}
favour SSA, while other popular representations within the functional programming
community are the A-normal form~\cite{ANF}, or Moggi's monadic
denotation~\cite{Monads}.

But there have been robust defences~\cite{CCExt} made for CPS in modern times and
our own decision to follow the CPS approach was based on an early commitment to free
ProofScript from stack overflow and to allow for the later testing of control flow
constructs that may benefit from a stackless runtime.

\section{CPS Generation}
\label{sec:org3a65eeb}

The traditional means of producing CPS is by a transformation from one lambda term to
another, but we get straight to it: our CPS terms are generated directly from a
monadic combinator language, whose types bear a clear resemblence to the encoding of
effects via monads, with the addition of a call-with-current-continuation primitive.

A motivating observation is that we can represent a partially evaluated CPS program
by our current continuation together with the name of the last returned value. Thus,
in the CPS term

\begin{displaymath}
  \lambda k. g\ y\ (\lambda r. f\ x\ (\lambda r'. r'\ r\ k))
\end{displaymath} 

we could represent the situation just after evaluating \(g\ y\) by a pair whose first
component is the bound variable representing the input to the current continuation,
and whose second component is a lambda term with a hole for that same continuation:

\begin{displaymath}
  (r, \lambda k. g\ y\ (\lambda r. ...))
\end{displaymath} 

The hole indicates where the computation continues, while the variable \(r\) names the
value of the computation thus far.

\subsection{Lambda terms}
\label{sec:org5458068}

The actual lambda calculus we use for representing ProofScript code, missing only the
ability to define functions by mutual recursion, can be given by the datatype:

\begin{small}
\begin{Verbatim}[commandchars=\\\{\}]
  \PY{k+kr}{data} \PY{k+kt}{Var} \PY{n}{a} \PY{o+ow}{=} \PY{k+kt}{Z}
             \PY{o}{|} \PY{k+kt}{S} \PY{n}{a}

  \PY{k+kr}{data} \PY{k+kt}{Term} \PY{n}{t} \PY{n}{a} \PY{o+ow}{=} \PY{k+kt}{V} \PY{n}{a}
                \PY{o}{|} \PY{k+kt}{App} \PY{p}{(}\PY{k+kt}{Term} \PY{n}{t} \PY{n}{a}\PY{p}{)} \PY{p}{(}\PY{k+kt}{Term} \PY{n}{t} \PY{n}{a}\PY{p}{)}
                \PY{o}{|} \PY{k+kt}{Abs} \PY{p}{(}\PY{k+kt}{Term} \PY{n}{t} \PY{p}{(}\PY{k+kt}{Var} \PY{n}{a}\PY{p}{)}\PY{p}{)}
                \PY{o}{|} \PY{k+kt}{Fix} \PY{p}{(}\PY{k+kt}{Term} \PY{n}{t} \PY{p}{(}\PY{k+kt}{Var} \PY{n}{a}\PY{p}{)}\PY{p}{)}
                \PY{o}{|} \PY{k+kt}{Prim} \PY{p}{(}\PY{n}{t} \PY{p}{(}\PY{k+kt}{Term} \PY{n}{t} \PY{n}{a}\PY{p}{)}\PY{p}{)}
\end{Verbatim}
\end{small}

That is, a term is either:

\begin{itemize}
\item a free variable drawn from the alphabet of frees \texttt{a}
\item an application
\item an abstraction
\item a fixpoint of an abstraction
\item some primitive structure \texttt{t} over terms.
\end{itemize}

For bound variables, we have used a trick due to Bird and Paterson~\cite{DeBruijnNested}, where we use
DeBruijn indices in a typesafe way. Abstractions bind over the first DeBruijn index
\texttt{Z}, with outer variables appearing as successors \texttt{S}. This basic idea has recently
been used to implement the Ermine compiler~\cite{Ermine}, and the inherent type safety it affords
should alleviate concerns about inadvertent variable capture arising during CPS
generation.

\subsection{CPS primitives}
\label{sec:orgc5663cd}

Our \texttt{CPS} type is to be a monad whose values mimic computation in our imperative
source language. Thus, while values in this type cumulatively assemble a lambda term
in CPS form, they also carry around an environment of variables that are in scope in
the source language.

There is also some book-keeping. CPS transformation introduces a lot of new bound
variables for continuations and intermediate values, and we want these to be hidden
from the user of our \texttt{CPS} type. Thus, we introduce a type

\begin{small}
\begin{Verbatim}[commandchars=\\\{\}]
  \PY{k+kr}{data} \PY{k+kt}{Val} \PY{o+ow}{=} \PY{k+kt}{Inter} \PY{k+kt}{Int} \PY{o}{|} \PY{k+kt}{Free} \PY{k+kt}{String}
\end{Verbatim}
\end{small}
consisting of indexed intermediate variables generated by CPS by an internal counter
and hidden from the user, as well as string variables that the user can generate for
themselves:

\begin{small}
\begin{Verbatim}[commandchars=\\\{\}]
  \PY{n}{nextIndex} \PY{o+ow}{::} \PY{k+kt}{CPS} \PY{k+kt}{Int}
  \PY{n}{free} \PY{o+ow}{::} \PY{k+kt}{String} \PY{o+ow}{\PYZhy{}\PYZgt{}} \PY{k+kt}{CPS} \PY{k+kt}{Val}
\end{Verbatim}
\end{small}

During CPS, we can emit more of the CPS'd term generated so far via the functions:

\begin{small}
\begin{Verbatim}[commandchars=\\\{\}]
  \PY{n}{nest} \PY{o+ow}{::} \PY{p}{(}\PY{k+kt}{Term} \PY{k+kt}{Exit} \PY{k+kt}{Val} \PY{o+ow}{\PYZhy{}\PYZgt{}} \PY{k+kt}{Term} \PY{k+kt}{Exit} \PY{k+kt}{Val}\PY{p}{)} 
          \PY{o+ow}{\PYZhy{}\PYZgt{}} \PY{k+kt}{CPS} \PY{n+nb}{()}
  \PY{n}{end} \PY{o+ow}{::} \PY{k+kt}{Term} \PY{k+kt}{Exit} \PY{k+kt}{Val} \PY{o+ow}{\PYZhy{}\PYZgt{}} \PY{k+kt}{CPS} \PY{k+kt}{Void}
\end{Verbatim}
\end{small}

The function \texttt{nest} builds more computation, while \texttt{end} terminates computation,
commiting a final and definitive lambda term as the result. As such, the use of the
uninhabited type \texttt{Void} indicates that the computation has no possible return value.

Values of type \texttt{CPS Val}, on the other hand, indicate that the computation represents
an expression. This refers to the situation described in the previous section: we
are holding onto a pair consisting of a lambda term with a hole together with a free
variable or the innermost intermediate value bound in a continuation.

\section{Pure Structured Programming}
\label{sec:org5a59ce0}

The building blocks for writing our pure structured programs can be likened to nodes
in the control flow graph of an SSA program. In SSA, assignments within a node are
translated to bindings of a right-hand side to a fresh variable. Thus, code such as

\begin{small}
\begin{Verbatim}[commandchars=\\\{\}]
  val x = 0
  x = x + 1
  x = x + 1
\end{Verbatim}
\end{small}
might become

\begin{small}
\begin{Verbatim}[commandchars=\\\{\}]
  val x = 0
  val x1 = x + 1
  val x2 = x1 + 1
\end{Verbatim}
\end{small}

The variable corresponding to the final assignment to \texttt{x}, namely \texttt{x2}, is then
exported from the node in the control flow graph and may be imported into successor
nodes by a special function \(\phi\).

In our CPS version, the successor nodes are continuations. Assignments are still
translated to rebindings, but the values after the final assignments are thrown as
extra arguments to these continuations.

Our \texttt{CPS} language structures programs into blocks, similar to nodes in a control
flow graph. A block is a function from a \texttt{CPS} computation to another \texttt{CPS}
computation. The block computes the set of possible assignments in its input
computation, and then ensures that any exit from this block via a continuation passes
on the latest values after assignment.

To support this, we add a primitive to our term language:

\begin{small}
\begin{Verbatim}[commandchars=\\\{\}]
  \PY{k+kr}{data} \PY{k+kt}{Exit} \PY{n}{a} \PY{o+ow}{=} \PY{k+kt}{Exit} \PY{k+kt}{Int} \PY{p}{(}\PY{p}{[}\PY{k+kt}{String}\PY{p}{]} \PY{o+ow}{\PYZhy{}\PYZgt{}} \PY{n}{a}\PY{p}{)}
  \PY{k+kr}{type} \PY{k+kt}{CPSTerm} \PY{n}{a} \PY{o+ow}{=} \PY{k+kt}{Term} \PY{k+kt}{Exit} \PY{n}{a}
\end{Verbatim}
\end{small}

The \texttt{Exit} primitive is a hole with an index that is matched up to an intermediate
continuation variable. The hole is filled based on the assignments to
variables given by the \texttt{[String]} argument, which are computed once those assignments
are fully known. It is an additional job of the \texttt{CPS} type to record assignments as
they occur in computations within a block, so that they can be later applied to the
appropriate \texttt{Exit} primitives.

\subsection{Call/CC and Loop}
\label{sec:orgb7b3343}

The three basic building blocks from which we build all others are two versions of
call-with-current-continuation and \texttt{loop}:

\begin{small}
\begin{Verbatim}[commandchars=\\\{\}]
  \PY{n}{callCC\PYZus{}} \PY{o+ow}{::} \PY{p}{(}\PY{k+kt}{CPS} \PY{k+kt}{Void} \PY{o+ow}{\PYZhy{}\PYZgt{}} \PY{k+kt}{CPS} \PY{k+kt}{Void}\PY{p}{)} 
             \PY{o+ow}{\PYZhy{}\PYZgt{}} \PY{k+kt}{CPS} \PY{n+nb}{()}
  \PY{n}{callCC} \PY{o+ow}{::} \PY{p}{(}\PY{p}{(}\PY{k+kt}{CPS} \PY{k+kt}{Val} \PY{o+ow}{\PYZhy{}\PYZgt{}} \PY{k+kt}{CPS} \PY{k+kt}{Void}\PY{p}{)} \PY{o+ow}{\PYZhy{}\PYZgt{}} \PY{k+kt}{CPS} \PY{k+kt}{Void}\PY{p}{)}
            \PY{o+ow}{\PYZhy{}\PYZgt{}} \PY{k+kt}{CPS} \PY{k+kt}{Val}
  \PY{n}{loop} \PY{o+ow}{::} \PY{p}{(}\PY{k+kt}{CPS} \PY{k+kt}{Void} \PY{o+ow}{\PYZhy{}\PYZgt{}} \PY{k+kt}{CPS} \PY{k+kt}{Void}\PY{p}{)} \PY{o+ow}{\PYZhy{}\PYZgt{}} \PY{k+kt}{CPS} \PY{k+kt}{Void}
\end{Verbatim}
\end{small}

Each of these functions:

\begin{enumerate}
\item inputs a block;
\item generates its CPS term;
\item computes its assignments;
\item nests the CPS'ed term inside its successor block.
\end{enumerate}

Specifically, \texttt{callCC\_} inputs a block which depends on an exit into the successor
block; the function \texttt{callCC} inputs a block which depends on an exit which
additionally receives a value to throw to the next block. Finally, \texttt{loop} inputs a
block which depends on an exit into itself.

For example, if \texttt{f} takes an exit \(k\) and produces a CPS'd term \(\phi(k)\) with
assignments to variables \(x\) and \(y\), then \texttt{callCC\_ f} will produce a CPS'd term

\begin{displaymath}
  (\lambda k. \phi(k\ x\ y))\ (\lambda x\ y. ...)  
\end{displaymath}
while \texttt{loop f} will produce a CPS'd term

\begin{displaymath}
  \texttt{fix}\ (\lambda\ loop\ x\ y. \phi(loop\ x\ y)).
\end{displaymath}

\subsection{Binding and assignment}
\label{sec:orge0f6701}

Assignments can only be made if we can statically determine how they should be
propagated to successor blocks. In the original concept of ProofScript, this was
controlled with the introduction of ``linear scopes". 

For our \texttt{CPS} type, the constraints are enforced mostly be the use of our basic
building blocks defined in the previous section, with one additional constraint: the
bodies of abstractions, effectively being dynamic nodes in the control flow, are
outright not allowed to assign to variables defined outside their scope. This is
enforced by ensuring that every \texttt{CPS} value tracks an environment of assignable
variables, which can be cleared whenever we compute the CPS of an abstraction.

\begin{small}
\begin{Verbatim}[commandchars=\\\{\}]
  \PY{n}{val} \PY{o+ow}{::} \PY{k+kt}{String} \PY{o+ow}{\PYZhy{}\PYZgt{}} \PY{k+kt}{CPS} \PY{k+kt}{Val} \PY{o+ow}{\PYZhy{}\PYZgt{}} \PY{k+kt}{CPS} \PY{k+kt}{Val} \PY{o+ow}{\PYZhy{}\PYZgt{}} \PY{k+kt}{CPS} \PY{k+kt}{Val}
  \PY{n}{assign} \PY{o+ow}{::} \PY{k+kt}{String} \PY{o+ow}{\PYZhy{}\PYZgt{}} \PY{k+kt}{CPS} \PY{k+kt}{Val} \PY{o+ow}{\PYZhy{}\PYZgt{}} \PY{k+kt}{CPS} \PY{n+nb}{()}
  \PY{n}{abs} \PY{o+ow}{::} \PY{k+kt}{String} \PY{o+ow}{\PYZhy{}\PYZgt{}} \PY{k+kt}{CPS} \PY{k+kt}{Val} \PY{o+ow}{\PYZhy{}\PYZgt{}} \PY{k+kt}{CPS} \PY{k+kt}{Val}
\end{Verbatim}
\end{small}

The expression \texttt{val x rhs body} evaluates \texttt{rhs} and then evaluates \texttt{body} in an
environment which contains a new assignable binding of \texttt{x} using the function
\texttt{withLocal}. The implementation makes use of a number of functions defined so far:

\begin{small}
\begin{Verbatim}[commandchars=\\\{\}]
  \PY{n}{val} \PY{o+ow}{::} \PY{k+kt}{String} \PY{o+ow}{\PYZhy{}\PYZgt{}} \PY{k+kt}{CPS} \PY{k+kt}{Val} \PY{o+ow}{\PYZhy{}\PYZgt{}} \PY{k+kt}{CPS} \PY{k+kt}{Val} \PY{o+ow}{\PYZhy{}\PYZgt{}} \PY{k+kt}{CPS} \PY{k+kt}{Val}
  \PY{n}{val} \PY{n}{x} \PY{n}{rhs} \PY{n}{body} \PY{o+ow}{=} \PY{n}{callCC} \PY{o}{\PYZdl{}} \PY{n+nf}{\PYZbs{}}\PY{n}{k} \PY{o+ow}{\PYZhy{}\PYZgt{}} \PY{k+kr}{do}
    \PY{n}{rhs\PYZsq{}} \PY{o+ow}{\PYZlt{}\PYZhy{}} \PY{n}{rhs}
    \PY{n}{nest} \PY{p}{(}\PY{n+nf}{\PYZbs{}}\PY{n}{inner} \PY{o+ow}{\PYZhy{}\PYZgt{}}
            \PY{k+kt}{App} \PY{p}{(}\PY{k+kt}{Abs} \PY{p}{(}\PY{k+kt}{Bnd} \PY{n}{x}\PY{p}{)}
                 \PY{p}{(}\PY{n}{abstract} \PY{p}{(}\PY{o}{==} \PY{k+kt}{Free} \PY{n}{x}\PY{p}{)} \PY{n}{inner}\PY{p}{)}\PY{p}{)}
            \PY{p}{(}\PY{k+kt}{V} \PY{n}{rhs\PYZsq{}}\PY{p}{)}\PY{p}{)}
    \PY{n}{withLocal} \PY{n}{x} \PY{n}{body}
    \PY{n}{k} \PY{n}{body}
\end{Verbatim}
\end{small}

The function \texttt{assign} is the basic primitive \emph{statement}, as indicated by its result
type \texttt{CPS ()}. Its definition checks whether the variable is assignable, establishes
a new binding for the variable, and registers a new assignment with a function
\texttt{tell}:

\begin{small}
\begin{Verbatim}[commandchars=\\\{\}]
  \PY{n}{assign} \PY{o+ow}{::} \PY{k+kt}{String} \PY{o+ow}{\PYZhy{}\PYZgt{}} \PY{k+kt}{CPS} \PY{k+kt}{Val} \PY{o+ow}{\PYZhy{}\PYZgt{}} \PY{k+kt}{CPS} \PY{n+nb}{()}
  \PY{n}{assign} \PY{n}{x} \PY{n}{rhs} \PY{o+ow}{=} \PY{k+kr}{do}
    \PY{n}{inScope} \PY{o+ow}{\PYZlt{}\PYZhy{}} \PY{n}{flip} \PY{n}{elem} \PY{o}{\PYZlt{}\PYZdl{}\PYZgt{}} \PY{n}{ask}
    \PY{k+kr}{if} \PY{n}{inScope} \PY{n}{x} \PY{k+kr}{then} \PY{k+kr}{do}
      \PY{n}{rhs\PYZsq{}} \PY{o+ow}{\PYZlt{}\PYZhy{}} \PY{n}{rhs}
      \PY{n}{nest} \PY{p}{(}\PY{n+nf}{\PYZbs{}}\PY{n}{inner} \PY{o+ow}{\PYZhy{}\PYZgt{}}
              \PY{k+kt}{App} \PY{p}{(}\PY{k+kt}{Abs} \PY{p}{(}\PY{k+kt}{Bnd} \PY{n}{x}\PY{p}{)}
                   \PY{p}{(}\PY{n}{abstract} \PY{p}{(}\PY{o}{==} \PY{k+kt}{Free} \PY{n}{x}\PY{p}{)} \PY{n}{inner}\PY{p}{)}\PY{p}{)}
              \PY{p}{(}\PY{n}{pure} \PY{n}{rhs\PYZsq{}}\PY{p}{)}\PY{p}{)}
      \PY{n}{tell} \PY{p}{(}\PY{n}{mempty}\PY{p}{,} \PY{p}{[}\PY{n}{x}\PY{p}{]}\PY{p}{)}
      \PY{k+kr}{else} \PY{n+ne}{error} \PY{p}{(}\PY{n}{x} \PY{o}{++} \PY{l+s}{\PYZdq{}}\PY{l+s}{ cannot be assigned here.}\PY{l+s}{\PYZdq{}}\PY{p}{)}
\end{Verbatim}
\end{small}

\subsection{Example}
\label{sec:orgc4de70d}

The following code is taken from the original language specification for "Babel-17",
a language developed by Obua~\cite{Babel17}:

\begin{small}
\begin{Verbatim}[commandchars=\\\{\}]
  a =\PYZgt{} b =\PYZgt{}
    if a == 0 then
      b
    else
      val a = a
        while b != 0 do
          if a \PYZgt{} b then
             a = a \PYZhy{} b
          else b = b \PYZhy{} a
        end
      end
    a
  end
\end{Verbatim}
\end{small}

To show how this code can be expressed by a \texttt{CPS} value, we'll first assume we have
the following primitive functions to hand:

\begin{small}
\begin{Verbatim}[commandchars=\\\{\}]
  \PY{n}{eq} \PY{o+ow}{::} \PY{k+kt}{CPS} \PY{k+kt}{Val} \PY{o+ow}{\PYZhy{}\PYZgt{}} \PY{k+kt}{CPS} \PY{k+kt}{Val} \PY{o+ow}{\PYZhy{}\PYZgt{}} \PY{k+kt}{CPS} \PY{k+kt}{Val}
  \PY{n}{gq} \PY{o+ow}{::} \PY{k+kt}{CPS} \PY{k+kt}{Val} \PY{o+ow}{\PYZhy{}\PYZgt{}} \PY{k+kt}{CPS} \PY{k+kt}{Val} \PY{o+ow}{\PYZhy{}\PYZgt{}} \PY{k+kt}{CPS} \PY{k+kt}{Val}
  \PY{n}{sub} \PY{o+ow}{::} \PY{k+kt}{CPS} \PY{k+kt}{Val} \PY{o+ow}{\PYZhy{}\PYZgt{}} \PY{k+kt}{CPS} \PY{k+kt}{Val} \PY{o+ow}{\PYZhy{}\PYZgt{}} \PY{k+kt}{CPS} \PY{k+kt}{Val}
  \PY{n}{not} \PY{o+ow}{::} \PY{k+kt}{CPS} \PY{k+kt}{Val} \PY{o+ow}{\PYZhy{}\PYZgt{}} \PY{k+kt}{CPS} \PY{k+kt}{Val}
\end{Verbatim}
\end{small}
as well as the following control structures

\begin{small}
\begin{Verbatim}[commandchars=\\\{\}]
  \PY{n}{cond} \PY{o+ow}{::} \PY{k+kt}{CPS} \PY{k+kt}{Val} \PY{o+ow}{\PYZhy{}\PYZgt{}} \PY{k+kt}{CPS} \PY{k+kt}{Val} \PY{o+ow}{\PYZhy{}\PYZgt{}} \PY{k+kt}{CPS} \PY{k+kt}{Val} \PY{o+ow}{\PYZhy{}\PYZgt{}} \PY{k+kt}{CPS} \PY{k+kt}{Val}
  \PY{n}{cond\PYZus{}} \PY{o+ow}{::} \PY{k+kt}{CPS} \PY{k+kt}{Val} \PY{o+ow}{\PYZhy{}\PYZgt{}} \PY{k+kt}{CPS} \PY{n+nb}{()} \PY{o+ow}{\PYZhy{}\PYZgt{}} \PY{k+kt}{CPS} \PY{n+nb}{()} \PY{o+ow}{\PYZhy{}\PYZgt{}} \PY{k+kt}{CPS} \PY{n+nb}{()}
\end{Verbatim}
\end{small}

We can then define a \texttt{while} loop in terms of \texttt{callCC\_} and \texttt{loop}:

\begin{small}
\begin{Verbatim}[commandchars=\\\{\}]
  \PY{n}{while} \PY{o+ow}{::} \PY{k+kt}{CPS} \PY{k+kt}{Val} \PY{o+ow}{\PYZhy{}\PYZgt{}} \PY{k+kt}{CPS} \PY{n}{a} \PY{o+ow}{\PYZhy{}\PYZgt{}} \PY{k+kt}{CPS} \PY{n+nb}{()}
  \PY{n}{while} \PY{n}{b} \PY{n}{body} \PY{o+ow}{=}
    \PY{n}{callCC\PYZus{}} \PY{p}{(}\PY{n+nf}{\PYZbs{}}\PY{n}{break} \PY{o+ow}{\PYZhy{}\PYZgt{}}
               \PY{n}{loop} \PY{p}{(}\PY{n+nf}{\PYZbs{}}\PY{n}{cont} \PY{o+ow}{\PYZhy{}\PYZgt{}}
                       \PY{n}{cond} \PY{n}{b} \PY{p}{(}\PY{n}{body} \PY{o}{\PYZgt{}\PYZgt{}} \PY{n}{cont}\PY{p}{)} \PY{n}{break}\PY{p}{)}\PY{p}{)}
\end{Verbatim}
\end{small}

And thus we can write:

\begin{small}
\begin{Verbatim}[commandchars=\\\{\}]
  \PY{n}{abs} \PY{l+s}{\PYZdq{}}\PY{l+s}{a}\PY{l+s}{\PYZdq{}} \PY{o}{\PYZdl{}} \PY{n}{abs} \PY{l+s}{\PYZdq{}}\PY{l+s}{b}\PY{l+s}{\PYZdq{}} \PY{o}{\PYZdl{}}
    \PY{n}{cond} \PY{p}{(}\PY{n}{eq} \PY{l+s}{\PYZdq{}}\PY{l+s}{a}\PY{l+s}{\PYZdq{}} \PY{l+s}{\PYZdq{}}\PY{l+s}{0}\PY{l+s}{\PYZdq{}}\PY{p}{)}
          \PY{l+s}{\PYZdq{}}\PY{l+s}{b}\PY{l+s}{\PYZdq{}}
          \PY{p}{(}\PY{k+kr}{do} \PY{n}{val} \PY{l+s}{\PYZdq{}}\PY{l+s}{a}\PY{l+s}{\PYZdq{}} \PY{l+s}{\PYZdq{}}\PY{l+s}{a}\PY{l+s}{\PYZdq{}} \PY{o}{\PYZdl{}} \PY{k+kr}{do}
                \PY{n}{while} \PY{p}{(}\PY{n}{not} \PY{p}{(}\PY{n}{eq} \PY{l+s}{\PYZdq{}}\PY{l+s}{b}\PY{l+s}{\PYZdq{}} \PY{l+s}{\PYZdq{}}\PY{l+s}{0}\PY{l+s}{\PYZdq{}}\PY{p}{)}\PY{p}{)} \PY{o}{\PYZdl{}} \PY{k+kr}{do}
                  \PY{n}{cond\PYZus{}} \PY{p}{(}\PY{n}{gq} \PY{l+s}{\PYZdq{}}\PY{l+s}{a}\PY{l+s}{\PYZdq{}} \PY{l+s}{\PYZdq{}}\PY{l+s}{b}\PY{l+s}{\PYZdq{}}\PY{p}{)}
                        \PY{p}{(}\PY{n}{assign} \PY{l+s}{\PYZdq{}}\PY{l+s}{a}\PY{l+s}{\PYZdq{}} \PY{p}{(}\PY{n}{sub} \PY{l+s}{\PYZdq{}}\PY{l+s}{a}\PY{l+s}{\PYZdq{}} \PY{l+s}{\PYZdq{}}\PY{l+s}{b}\PY{l+s}{\PYZdq{}}\PY{p}{)}\PY{p}{)}
                        \PY{p}{(}\PY{n}{assign} \PY{l+s}{\PYZdq{}}\PY{l+s}{b}\PY{l+s}{\PYZdq{}} \PY{p}{(}\PY{n}{sub} \PY{l+s}{\PYZdq{}}\PY{l+s}{b}\PY{l+s}{\PYZdq{}} \PY{l+s}{\PYZdq{}}\PY{l+s}{a}\PY{l+s}{\PYZdq{}}\PY{p}{)}\PY{p}{)}
                \PY{l+s}{\PYZdq{}}\PY{l+s}{a}\PY{l+s}{\PYZdq{}}\PY{p}{)}
\end{Verbatim}
\end{small}

Notice how similar this code is to the source language, despite being shallow
embedded in another functional language. The combinators generate a single pure
lambda term, where assignments are replaced by variable rebinding and their final
values propagated through continuations. In fact, after beta-eta reduction, we obtain
the concise and pure lambda term:

\begin{align*}
  \lambda i\ k. k\ (\lambda j\ k'. &if\ i = 0:\\
  &\quad k'\ j\\
  &else:\\
  &\quad\quad fix\ (\lambda loop\ a\ b.\\
  &\quad\quad\quad if\ b \neq 0:\\
  &\quad\quad\quad\quad if\ a > b:\\
  &\quad\quad\quad\quad\quad loop\ (a - b)\ b\\
  &\quad\quad\quad\quad else:\\
  &\quad\quad\quad\quad\quad loop\ a\ (b - a)\\
  &\quad\quad\quad else\ k'\ a)\ i\ j)
\end{align*}

\section{Conclusion}
\label{sec:orge5d9e76}

As we have shown, a pure functional language which supports assignment and structured
programming constructs can be embedded as a datatype \texttt{CPS} whose combination closely
resembles our source language \emph{ProofScript}. As such, the datatype \texttt{CPS} can be used
to express the semantics of ProofScript, as well as acting as the first compilation
phase in our production compiler. The datatype ensures a certain amount of
type-safety, distinguishing as it does between expressions of type \texttt{CPS Val},
statements of type \texttt{CPS ()} and computations that have exited via a continuation of
type \texttt{CPS Void}.

The basic ingredients here are a function to implement two basic control flow
mechanisms: looping and call-with-current-continuation. From these, a variety of
other control flow constructs can be obtained. The coordination of assignments
through these is automatically and safely handled by the basic datatype, leaving us
working with what feels like a pure but structured imperative language. 

In further work, we would like to formally investigate and explore our intuition that
the resulting language and the way it structures control flow closely resembles
static-single assignment.

\bibliographystyle{plain}
\bibliography{proofpeer}{}
\end{document}